\documentclass{article}

\usepackage[nonatbib, preprint]{neurips_2022}

\usepackage[utf8]{inputenc} 
\usepackage[T1]{fontenc}    
\usepackage{hyperref}       
\usepackage{url}            
\usepackage{booktabs}       
\usepackage{amsfonts}       
\usepackage{nicefrac}       
\usepackage{microtype}      
\usepackage{xcolor}         %
\usepackage{graphicx}

\usepackage{natbib}
\title{WeatherBench Probability: A benchmark dataset for probabilistic medium-range weather forecasting along with deep learning baseline models}

\author{%
  Sagar Garg \\
  Technical University of Munich, Germany
   \And
   Stephan Rasp \\
   ClimateAi, Inc \\
   \texttt{raspstephan@gmail.com} \\
   \AND
   Nils Thuerey \\
   Technical University of Munich, Germany \\
}

\begin{document}

\maketitle

\begin{abstract}
WeatherBench is a benchmark dataset for medium-range weather forecasting of geopotential, temperature and precipitation, consisting of preprocessed data, predefined evaluation metrics and a number of baseline models. WeatherBench Probability extends this to probabilistic forecasting by adding a set of established probabilistic verification metrics (continuous ranked probability score, spread-skill ratio and rank histograms) and a state-of-the-art operational baseline using the ECWMF IFS ensemble forecast. In addition, we test three different probabilistic machine learning methods---Monte Carlo dropout, parametric prediction and categorical prediction, in which the probability distribution is discretized. We find that plain Monte Carlo dropout severely underestimates uncertainty. The parametric and categorical models both produce fairly reliable forecasts of similar quality. The parametric models have fewer degrees of freedom while the categorical model is more flexible when it comes to predicting non-Gaussian distributions. None of the models are able to match the skill of the operational IFS model. We hope that this benchmark will enable other researchers to evaluate their probabilistic approaches.
\end{abstract}

\section{Introduction}
WeatherBench \citep{Rasp2020c} introduced a benchmark dataset for medium-range, data-driven weather forecasting, focused on 3 and 5 day predictions of global geopotential, temperature and precipitation fields. The three core components of WeatherBench are 1) a preprocessed training and validation dataset hosted in a convenient format on a public server, 2) a set of precisely defined evaluation metrics and 3) a set of baselines, ranging from simple ones such as climatologies and persistence to dynamical models run at different resolutions and simple machine learning models. 

The idea behind WeatherBench is to provide a framework in which to evaluate progress in data-driven weather forecasting. Since its publication, WeatherBench has been used in several papers. \citet{Weyn2020a} built a convolutional neural network (CNN) trained using a cubed-sphere coordinate transformation. They then extended this approach for subseasonal prediction \citep{Weyn2021Sub-SeasonalModels}. \citet{Rasp2021} used a deep Resnet architecture pretrained on climate model simulations that achieves comparable skill to dynamical baselines at similar resolutions. \citet{Gneiting2019} used the WeatherBench dataset to compare different evaluation metrics. \citet{Keisler2022} trained a graph neural network and achieved impressive skill on metrics very similar to those suggested in WeatherBench. 

One short-coming of the original WeatherBench benchmark is that it exclusively focuses on deterministic prediction. However, in medium-range weather forecasting is was long ago recognized that the inherent unpredictability of the atmosphere necessitates a probabilistic approach \citep{Bauer2015}. For this reason, virtually all operational weather forecasting centers run ensemble forecasts, where each ensemble member represents one realistic potential future outcome. Here we aim to extend WeatherBench by adding a probabilistic dimension to the verification metrics and baselines. Building on this we then train a range of probabilistic neural networks. 

There is some previous work in building probabilistic data-driven methods. \citet{Scher2020} tested several methods to create a data-driven ensemble forecast: random initial condition perturbations, singular vector initial condition perturbations and different random seeds for the neural networks. The latter results in the best probabilistic scores. \citet{Clare2021} explored dropout ensembles, similar to this work. We will contrast our results to theirs below.

We begin by briefly summarizing the original WeatherBench data and describing the new evaluation metrics and baselines in Section \ref{sec:data}. Then, in Section \ref{sec:methods} we review relevant literature in probabilistic machine learning before describing the three methods we will use. This is followed by the results of our methods in Section \ref{sec:results}, which we discuss further in Section \ref{sec:discussion}, before summarizing our findings in Section \ref{sec:conclusion}.

\section{Data, Evaluation and Baselines}
\label{sec:data}

\subsection{Data}
Here we will provide a brief summary of the WeatherBench dataset. For a detailed description refer to \citet{Rasp2020c}. The latest version of the data is available at \url{https://github.com/pangeo-data/WeatherBench}. WeatherBench contains regridded ERA5 \citep{Hersbach2020} data from 1979 to 2018 at hourly resolution with 2017 and 2018 set aside for evaluation. Here we use the coarsest horizontal resolution, 5.625$^{\circ}$ (32 $\times$ 64 grid points), at 7 vertical levels (50, 250, 500, 600, 700, 850, and 925 hPa). 

As a probabilistic baseline we use the 50-member operational ECMWF IFS ensemble, archived in the THORPEX Interactive Grand Global Ensemble (TIGGE) archive \citep{Bougeault2010}. The target variables 500\,hPa geopotential (Z500), 850\,hPa temperature (T850), 2\,m temperature (T2M) and 6h accumulated precipitation (PR) for 2017 and 2018 regridded to match WeatherBench are now also available at \citet{Rasp2022}. 

\subsection{Evaluation}

For evaluating the probabilistic forecasts, we chose several well-established metrics. First, we evaluate the deterministic skill of the ensemble mean using the root mean squared error (RMSE) defined as 

\begin{equation}
\label{eq:rmse}
    \textrm{RMSE} = \frac{1}{N_{\textrm{forecasts}}} \sum_i^{{N_{\textrm{forecasts}}}} \sqrt{\frac{1}{N_{\textrm{lat}} N_{\textrm{lon}}} \sum_j^{N_{\textrm{lat}}} \sum_k^{N_{\textrm{lon}}} L(j) (\bar{f}_{i, j, k} - t_{i, j, k})^2} 
\end{equation}
where $\bar{f}$ is the ensemble mean forecast and $t$ is the ERA5 truth. $L(j)$ is the latitude weighting factor for the latitude at the $j$th latitude index:
\begin{equation}
    L(j) = \frac{\cos( \textrm{lat}(j))}{\frac{1}{N_{\textrm{lat}}} \sum_j^{N_{\textrm{lat}}} \cos( \textrm{lat}(j)) }
\end{equation}

As a first-order measure of the reliability of the ensemble, we look at the spread-skill ratio ($\textrm{Spread} / \textrm{RMSE}$), where $\textrm{Spread}$ defined as
\begin{equation}
\label{eq:spread}
    \textrm{Spread} = \frac{1}{N_{\textrm{forecasts}}} \sum_i^{{N_{\textrm{forecasts}}}} \sqrt{\frac{1}{N_{\textrm{lat}} N_{\textrm{lon}}} \sum_j^{N_{\textrm{lat}}} \sum_k^{N_{\textrm{lon}}} L(j) \textrm{var}(f_{j, k}) }
\end{equation}
where $\textrm{var}(f_{j, k})$ is the variance in the ensemble dimension. A perfectly reliable ensemble should have a spread-skill ratio of 1. Smaller values indicate underdispersion (i.e. the probabilistic forecast is overconfident in its forecasts), larger values overdispersion \citep{Fortin2014}.


As a more detailed visual evaluation of reliability we use rank histograms, described in detail in \citet{Wilks2006}(p. 316). A perfect rank histogram is uniform, while a U-shape indicates underdispersion and a reverse U-shape overdispersion.

Finally, calibration and sharpness of the ensemble together is evaluated using the continuous ranked probability score (CRPS) \citet{Wilks2006}(p. 302), defined as:

\begin{equation}
    \textrm{CRPS} = \int_{-\infty}^{\infty} [F(f) - 1(t \leq z)]^2 dz,
\end{equation}
where $F$ denotes the CDF of the forecast distribution and $1(t \leq z)$ is an indicator function that is 1 if $t \leq z$ and 0 otherwise. In the case of a deterministic forecast the CRPS reduces to the mean absolute error. For computing the CRPS for ensemble forecasts, we use the implementation in the \texttt{xskillscore}\footnote{\url{https://xskillscore.readthedocs.io}} Python package.

\section{Probabilistic neural networks}
\label{sec:methods}

We explore three different methods to create probabilistic predictions using neural networks. The basic network architecture of all networks is a deep Resnet, which mirrors that of \citet{Rasp2021} with only the final layer changed. The input variables are geopotential, temperature, zonal and meridional wind and specific humidity at seven vertical levels (50, 250, 500, 600, 700, 850 and 925hPa), 2-meter temperature, 6-hourly accumulated precipitation, the top-of-atmosphere incoming solar radiation, all at the current time step $t$, $t - 6h$ and $t - 12h$, and, finally three constant fields: the land-sea mask, orography and the latitude at each grid point. All variables, levels and time steps are stacked to create a 114 channel input. All features are normalized $(x \gets \frac{x-\mathrm{mean}}{\mathrm{std}})$, except precipitation where $\mathrm{mean}=0$ to keep values positive. Precipitation is also log-transformed $(\mathrm{PR} \gets \log(\mathrm{PR}+\epsilon)-\log(\epsilon))$, where $\epsilon=0.001$. For evaluation the transformation is reversed again. All models were trained using ERA5 data from 1979 to 2015, with 2016 used for validation and 2017/18 for testing. Predictions are made for a lead-time of 3 days.\footnote{Ideally we would have also trained models for 5 days lead time and continuous models. However, because the first two authors moved on to different positions shortly after the initial work was finished, this was not possible. We would expect qualitatively similar results for other lead times.} The training procedure is the same as in \citet{Rasp2021}. No pretraining with climate models was used for any of the experiments in this paper.

\subsection{Monte-Carlo Dropout (MC Dropout)}
Dropout \citep{Srivastava2014} was introduced as a simple way of combating overfitting in neural networks. Typically, dropout is only applied during training but then turned off during inference. In fact, our base architecture described above uses dropout with a rate of 0.1. However, it is also possible to use dropout during inference to obtain a stochastic prediction \citep{Gal2015}. Here we train our model with different dropout rates, 0, 0.1, 0.2 and 0.5, and create 50 random realizations for the test dataset to match the 50-member IFS ensemble.


\subsection{Parametric prediction}
Another way to create a probabilistic forecast using neural networks is to directly predict the parameters of a distribution. This is a common technique in ensemble post-processing. Here we follow the methodology of \citet{Rasp2018d}. The first step is to prescribe a distribution for the target variables. For Z500, T850 and T2M we use a Gaussian distribution parameterized by its mean $\mu$ and variance $\sigma$. This means that instead of a single output channel, the network now has two channels for $\mu$ and $\sigma$. To optimize for the distributional parameters we chose the CRPS as a probabilistic loss function. For a Gaussian distribution, there exists a closed-form, differentiable solution of the CRPS:
\begin{equation}
\mathrm{}{CRPS}\left(F_{\mu, \sigma}, y\right)= \sigma\left\{\frac{y-\mu}{\sigma}\left[2 \Phi\left(\frac{y-\mu}{\sigma}\right)-1\right]+2 \varphi\left(\frac{y-\mu}{\sigma}\right)-\frac{1}{\sqrt{\pi}}\right\}
\end{equation}

For precipitation, a Gaussian distribution would be a poor fit because precipitation tends to be very skewed and is censored at zero. Several distributions have been proposed in literature such as a left-censored generalized extreme value distribution \citep{Scheuerer2014a}. We struggled to implement these for neural network training. The closed-forms of these distributions are quite complicated with limited valid ranges for some parameters. Unfortunately, we did not end up figuring out how to achieve a stable training using these distributions. We will discuss this in Section \ref{sec:discussion}.

For evaluating the parametric forecasts we computed the CRPS and the spread-skill ratio directly from the predicted parametric distributions but created a randomly sampled 50-member ensemble for computing the rank histogram to match the other methods.

\subsection{Categorical prediction}
Finally, we test a non-parametric, probabilistic method based on discretized bins. This has recently been used by \citet{Sonderby2020} and \citet{Espeholt2021} for precipitation nowcasting. The idea is to divide the value range to be predicted into discrete bins and predict the probability for each of these bins. The problem is then simply a multi-class classification problem. The number of channels in the last layer are changed to match the number of bins, followed by a softmax layer. The targets are one-hot encoded, with a one indicating the correct bin and zeros otherwise. As a loss function we use the standard categorical cross-entropy, also called log-loss

There are some subtleties in how to construct the bins. Generally, there is a trade-off between the bin width, which can be thought of as the probabilistic resolution, and the number of bins. Higher resolution will lead to finer probability distributions but will also make training harder. Further, the residual blocks of the core architecture have 128 channels which sets an upper limit to the number of channels. However, taking temperature prediction as an example, temperature values can fluctuate by around 100C from the poles to the deserts, which means that the bins would be rather large. For this reason, for geopotential and temperature, we predict the change in time ($X_{t=3d} - X_{t=0}$; after normalization) rather than the absolute values. These differences will have a much smaller spread. Before evaluation, we then simply add the difference. 

Note that we train our model separately for each variable, unlike in the other experiments where Z, T, T2M are trained together and precipitation trained separately. Empirically, this has given better results compared to training all variables together. For Z, T and T2M, we use 50 bins ranging from -1.5 to 1.5. For precipitation, we use 50 bins from 0 to 8 (after the log-transform).

\section{Results}
\label{sec:results}

\begin{table}
\caption{RMSE, Spread-skill ratio and CRPS for different methods for 3 day forecasts.}
\label{tab:results}
\hspace{-50pt}
\resizebox{1.2\textwidth}{!}{%
\begin{tabular}{ccccc|cccc|cccc}
             & \multicolumn{4}{c}{RMSE of ensemble mean} & \multicolumn{4}{c}{Spread-Skill Ratio} & \multicolumn{4}{c}{CRPS}       \\ \hline \hline
             & Z500 (m$^2$s$^{-2}$)         & T850 (K)       & T2M (K)    & TP (mm)       & Z500        & T850 & T2M & TP       & Z500 (m$^2$s$^{-2}$)         & T850 (K)       & T2M (K)    & TP (mm)      \\ \hline
\begin{tabular}[c]{@{}c@{}}MC Dropout \\ (Dr=0.1)\end{tabular} & 312.96 & 1.80 & 1.52 & 2.2 & 0.40 & 0.34 & 0.39 & 0.13 & 155.70 & 1.03 & 0.77 & 0.57 \\
Parametric   & 315.30    & 1.82    & 1.55   & -        & 0.87     & 0.92    & 0.95    & -       & 142.67 & 0.90 & 0.70 & -       \\
Categorical  & 327.48    & 1.80    & 1.49   & 2.2   & 0.91     & 0.94    & 0.98    & 1.24    & 142.59 & 0.87 & 0.65 & 0.47 \\
Tigge (3/5 days)        & 145/297    & 1.20/1.73    & 1.26/1.57   & 2.02/2.15   & 1.05/1.00     & 0.93/0.96    & 0.69/0.80    & 0.84/0.85    & 65.6/127  & 0.60/0.83 & 0.58/0.70 & 0.41/0.47 \\
Determinstic & 313.70   & 1.79    & 1.53   & 2.2   & -        & -       & -       & -       & 194.90 & 1.24 & 0.96 & 0.65 \\
\citet{Clare2021}  & 375/627 & 2.11/2.91 & & &  & & & & 211/1500 & 1.22 / 1.69 & &  \\
\end{tabular}%
}
\end{table}

\begin{figure}

\includegraphics[width=\textwidth]{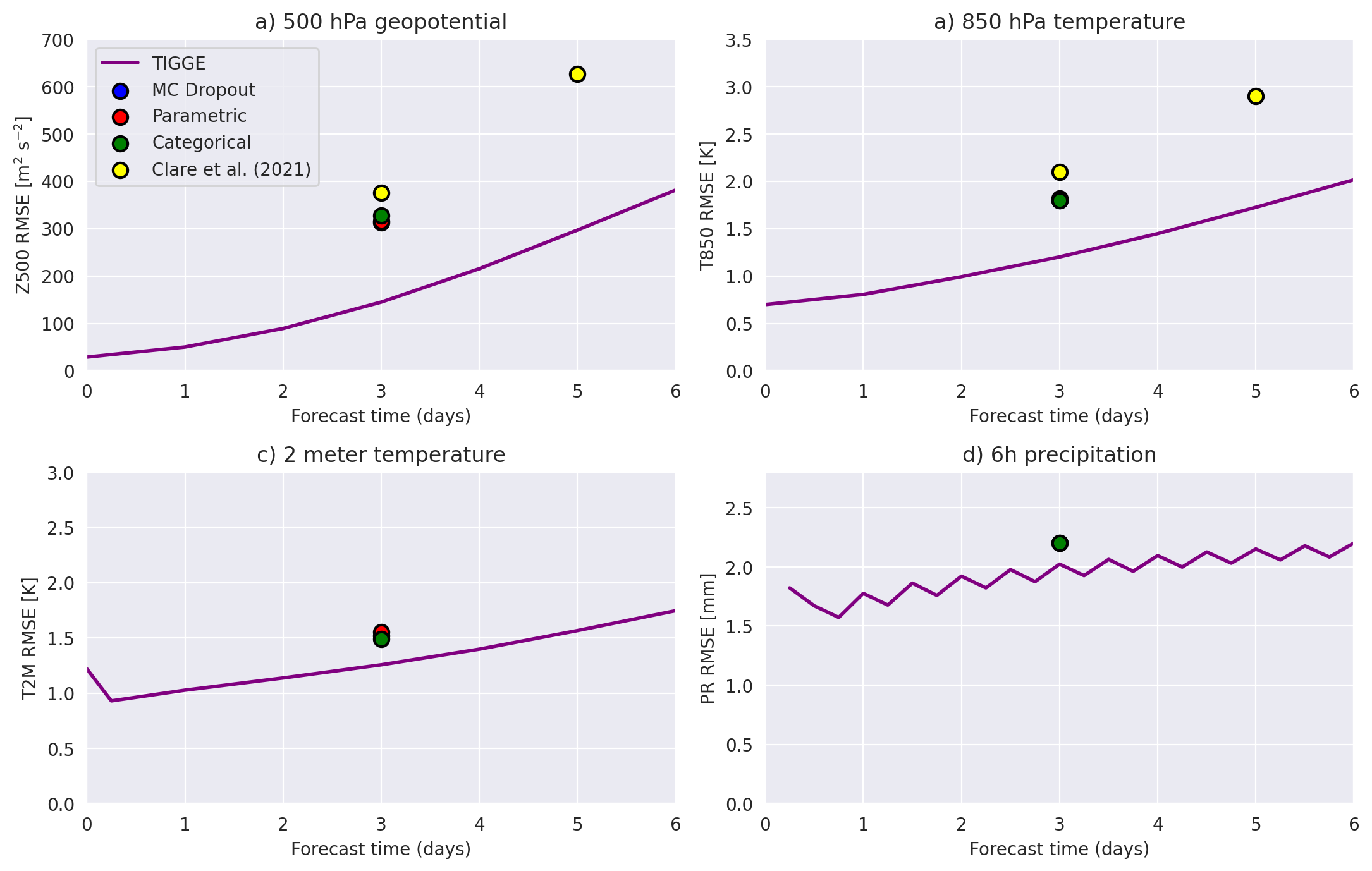}
\hrule
\includegraphics[width=\textwidth]{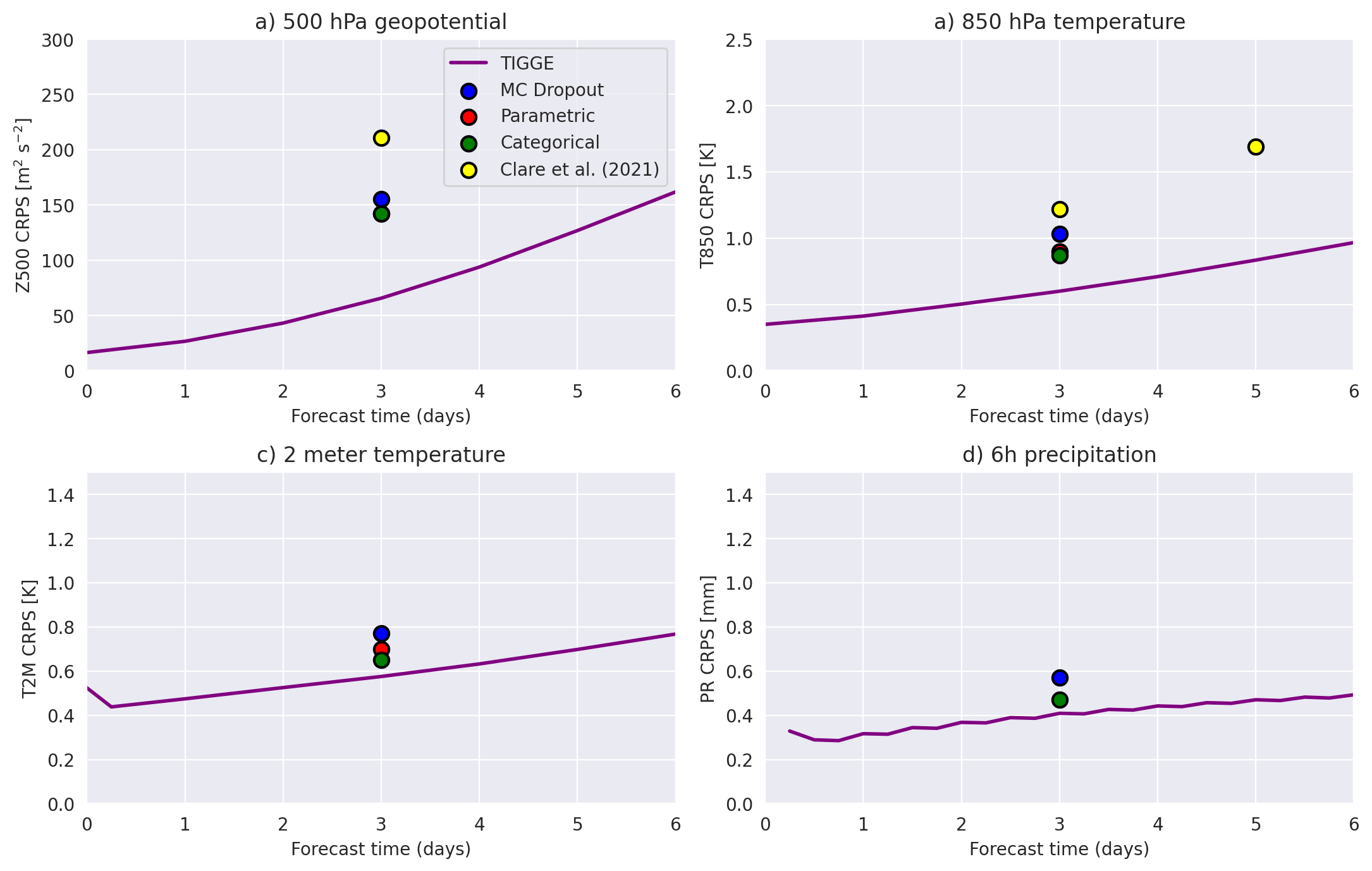}
\caption{(top) Ensemble mean RMSE for TIGGE baseline and different data-driven methods. (bottom) Same for CRPS.}
\label{fig:results}
\end{figure}

\begin{figure}

\includegraphics[width=\textwidth]{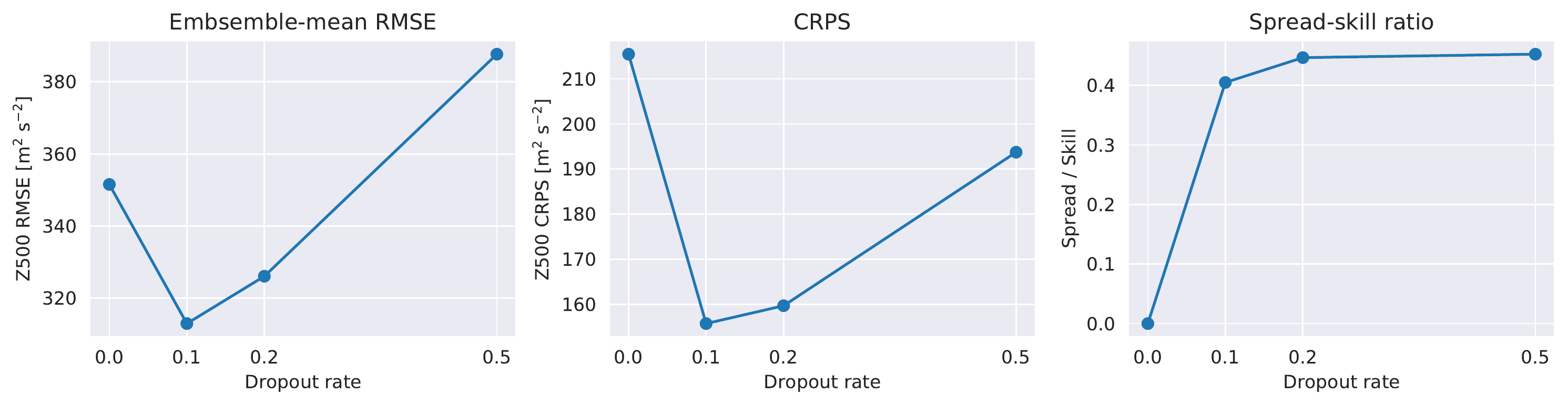}
\caption{RMSE, CRPS and spread-skill ratio for different dropout rates for Z500.}
\label{fig:dropout_sensitivity}
\end{figure}

The results for the different verification metrics are shown in Table~\ref{tab:results} and Fig.~\ref{fig:results}. For MC dropout we also tested the sensitivity to the dropout rate, shown in Fig.~\ref{fig:dropout_sensitivity} (only Z500 is shown; other variables behave qualitatively similarly). The ensemble mean RMSE and the CRPS is lowest for a dropout rate of 0.1. The spread-skill ratio shows that the dropout ensemble is severely underdispersive with the spread being less than half of what it should be. These results are consistent with \citet{Clare2021}. Increasing the dropout rate beyond 0.1 has a slight positive effect on the spread-skill ratio but this comes at the cost of severely decreasing skill measured by the ensemble mean RMSE and CRPS. This will be further discussed in Section \ref{sec:discussion}. For now, we use a dropout rate of 0.1 for all further evaluation.

The ensemble mean RMSE is reasonably similar between all deep learning methods with slight fluctuations between the variables. We also added the RMSE of the deterministic deep learning model without pretraining described in \citet{Rasp2021} as well as the scores reported in \citet{Clare2021}. The operational TIGGE ensemble outperforms the data-driven methods quite significantly. The only exception is precipitation, but, as already discussed in \citet{Rasp2021}, the RMSE is not a particularly fitting metric to evaluate a highly intermittent and skewed field like precipitation. 

The spread-skill ratio shows that MC dropout is severely underdispersive for all variables. The parametric and categorical networks have similar ratios just under one for geopotential and temperature, while the ratio is above one for categorical precipitation. TIGGE has ratios very close to one for Z500 and T850 but is somewhat underdispersive for T2M. It is important to remember here though that the TIGGE results are not post-processed, which would likely improve calibration. 

\begin{figure}

\includegraphics[width=\textwidth]{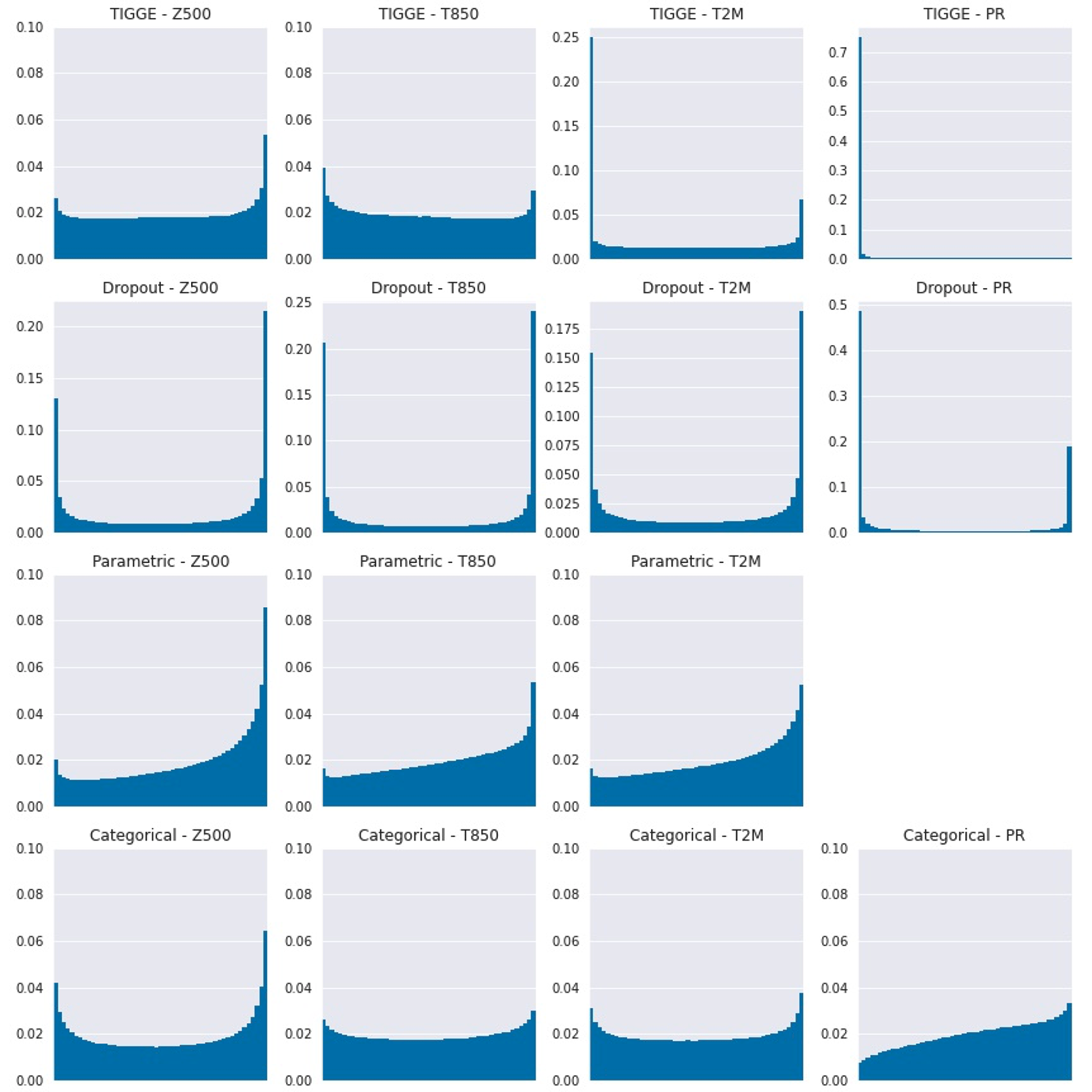}
\caption{Rank histograms for different variables and methods. All rank histograms have 51 bins.}
\label{fig:rank_histograms}
\end{figure}

Further insight into calibration and reliability can be gained from looking at the rank histograms in Fig.~\ref{fig:rank_histograms}. Here, we see a very strong U-shape for MC dropout. TIGGE only has a slight U-shape for Z500 and T850 but strong peaks on the left-hand side for T2M and PR. For T2M, this hints at a bias issue but would require further investigation. The peak for PR could be a manifestation of the well-known drizzle bias. The parametric predictions are skewed which is a sign of a high-bias. We did not further investigate the cause for this. The categorical predictions do not have a bias and are only moderately U-shaped, suggesting good calibration.

The CRPS evaluates both calibration and sharpness. Again we see similar scores for the parametric and categorical networks, with slightly better results for categorical. MC dropout is worse on CRPS, a consequence of its underdispersion. Further, we see that the deterministic forecast (for which the CRPS is simply the MAE) naturally scores worse since it does not provide any probabilistic information. Again, TIGGE has significantly better CRPS scores compared to the data driven methods.

\section{Discussion}
\label{sec:discussion}
\subsection{Verification caveats}
There are several aspects to consider in the above verification. First, the verification is done at a very coarse grid, 5.625 degrees. The operational IFS ensemble forecast runs at a resolution of around 13 km. The required regridding could potentially add a source of error, which might partly account for the initial error at $t=0h$ of the operational ensemble. However, another cause of initial error is the fact that the operational forecasts are started from different initial conditions, the operational analysis as opposed to the ERA5 reanalysis. For a fair comparison of pure forecasting skill, each method should be compared against its own analysis, starting from zero error at $t=0h$. 

For precipitation, the verification metrics used, RMSE and CRPS, are poor choices as they suffer heavily from the double penalty problem. In the original WeatherBench paper and here, we still decided to include the metrics but urge everyone to interpret these results with the appropriate caution.



\subsection{Parametric versus categorical prediction}
Both, parametric and categorical forecasts performed well in this study, with similar verification scores. They also both have been used successfully in previous research, even if parametric methods have a much longer history, especially in post-processing. Parametric approaches have the advantage of simplicity. Only a few (usually 2-3) parameters are to be estimated to produce a full distribution. For parameters where the underlying distribution is thought to be well known, e.g. temperature and geopotential, a parametric approach is a great choice. However, for more unusually distributed variables, e.g. precipitation, problems can arise. First, often the "true" distribution is not know which requires assumptions to be made. If those assumptions are not true, an artificial error is introduced. Second, more complex distribution functions and their derivatives required for gradient descent optimization can be difficult to implement. In the case of precipitation, we attempted to implement a censored extreme value distribution described in \citet{Scheuerer2014}. However, we encountered exploding gradients during optimization, likely a result of the limited valid ranges of some of the distributional parameters. One could likely fix this by adding constraints on the parameter values and tinkering with the implementation of the loss. However, it shows some of the difficulty with the parametric approach. 

The categorical approach, on the other hand, requires no assumptions on the distribution. In theory, any distribution can be learned. This is potentially a big advantage of this method, especially for more complex distributions. The trade-off is that instead of learning 2-3 values, now $>$100 values have to be estimated, which can be problematic for small sample sizes. In fact, we have observed that initially the estimated "distributions" are not smooth and only become so after training. Further, one introduces a probabilistic resolution, i.e. the bin width. As described in the Methods, this required a reframing of the problem for geopotential and temperature to predict the change rather than the absolute value. Nevertheless, the categorical approach, given enough samples, turns out to work very well empirically and is easy to implement. 

\subsection{Spatially and temporally correlated forecasts}
A final topic to discuss is spatial correlation in the forecast fields. The parametric and categorical methods predict separate distributions at each grid point and at each point in time. While this is fine for some applications, others also require knowledge about spatial and temporal patterns. For flood forecasting, for example, one is not simply interested in the distribution of rain at a single location in 5 days time but rather in the distribution of cumulative rain in the entire catchment area. For rain, which can have very different characteristics, e.g. convective versus stratiform, adding the distributions in space and time will not lead to the correct answer. Further, visual inspection of weather maps is still a very common use case for decision makers. With point-wise distribution methods, it is impossible to create "realistic" weather maps. Monte-Carlo dropout produces an ensemble of spatially coherent outputs but produces less reliable results compared to the other methods.

To solve this issue, generative methods could be an option. Generative adversarial networks (GANs) \citep{Goodfellow2014} are one way of producing realistic weather output. Recently this has been used for precipitation nowcasting \citep{Ravuri2021SkillfulRadar} and downscaling \citep{Price2022, Harris2022}. While GANs are, in principal, a well-suited method to produce coherent output, in practice they can be hard to train and empirically don't always produce the correct distribution.

\section{Conclusion}
\label{sec:conclusion}
Here we introduced WeatherBench Probability, a probabilistic extension to the WeatherBench benchmark dataset. We added commonly used probabilistic verification metrics along with an operational numerical weather prediction baseline. Further, we trained three different sets of deep learning models to produce probabilistic predictions for 3 days ahead, Monte-Carlo dropout, parametric and categorical networks. Our version of Monte-Carlo dropout produced severely underdispersive results. The other two methods produced similarly good results across all scores, with trade-offs between the two discussed in Section~\ref{sec:discussion}. 

We hope that this paper complements the WeatherBench dataset and enables fast and easy comparison between different data-driven forecasting methods. In the last year or two, a number of promising approaches have been proposed and others are in development which makes for an exciting future for this research field.

\bibliography{bibliography}







\end{document}